\input harvmac.tex
\Title{\vbox{\baselineskip12pt\hbox{IFT-UAM/CSIC-98-14}
\hbox{MRI/PHY/P980961}
\hbox{hep-th/9810140}}}
{\vbox{\centerline{Boundary Fluctuations of AdS String}}}
\centerline{{\bf Sudipta Mukherji${}^1$,  Sudhakar Panda}${}^2$}
\smallskip\centerline{\it ${}^1$Instituto de Fisica Te\'orica}
\smallskip\centerline{\it Universidad Aut\'onoma de Madrid, Madrid, Spain}
\smallskip\centerline{e-mail: mukherji@delta.ft.uam.es}
\vskip .2in
\smallskip\centerline{\it ${}^2$Mehta Research Institute}
\smallskip\centerline{\it of Mathematics and Mathematical Physics}
\smallskip\centerline{\it Allahabad, 211 019, India}
\smallskip\centerline{e-mail: panda@mri.ernet.in}

\vskip .3in
We analyse the world-sheet perturbations of string theory
formulated around $AdS_3$ background. We identify a set 
of operators that, while added to the world-sheet action,
generate the boundary fluctuations of $AdS_3$. 
The effect of these operators can be collectively defined
in terms of Liouville field living on the $AdS_3$ boundary.
We then study various deformations of $AdS_3$ generated 
by boundary fluctuations by turning on suitable world-sheet
couplings. We also discuss certain small fluctuations around 
the BTZ black hole.
\Date{10/98}

\lref\brown{J.D. Brown and M. Henneaux, {\it Comm. Math. Phys.}
{\bf 104} (1986) 207.}

\lref\strominger{ A. Strominger, {\it Black hole entropy from near 
horizon microstates}, hep-th/9712251.}

\lref\henneaux{O. Coussaert, M. Henneaux and P. van Driel,
{\it Class. Quant. Grav.} {\bf 12} (1995) 2961.}

\lref\btz{M. Ba\~nados, C. Teitelboim and J. Zanelli,
{\it Phys. Rev. Lett.} {\bf 69} (1992) 1849.}

\lref\seiberg{A. Giveon, D. Kutasov and N. Seiberg, {\it
Comments on string theory on $AdS_3$}, hep-th/9806194.}

\lref\navarro{J. Navarro-Salas and P. Navarro, {\it A note
on Einstein gravity on $AdS_3$ and boundary conformal
field theory}, hep-th/9807019.}

\lref\martinec{E. Martinec, {\it Conformal field theory, geometry, and 
entropy}, hep-th/9809021.}

\lref\mms{S. Mukherji, S. Mukhi and A. Sen, {\it Phys. Lett.}
{\bf B275} (1992) 39.}

\lref\bbg{K. Behrndt, I. Brunner and I. Gaida, {\it Entropy and 
conformal field theories of $AdS_3$ models}, hep-th/9804159.}

\lref\bss{D. Bermingham, I. Sachs and S. Sen, {\it Entropy
of three dimensional black holes in string theory}, hep-th/9801019.}

\lref\ms{J. Maldacena and A. Strominger, {\it $AdS_3$ black holes
and stringy exclusion principle}, hep-th/9804085.}

\lref\carlip{S. Carlip, {\it What we don't know about BTZ 
black hole entropy}, hep-th/9806026.}


In a three dimensional theory of gravity, the gravitational field contains  
no local 
degrees of freedom and therefore these theories are completely
characterized
by certain global set of data. In particular, for the case of
$AdS_3$, the global information is encoded in the space-time 
diffeomorphisms which survive at the two dimensional boundary of
$AdS_3$. It is, by now, well-known that such diffeomorphisms can 
be described by conformal field theory (CFT) living on the boundary
\refs{\brown}. Thus the asymptotic symmetry group of $AdS_3$ is 
generated by infinite dimensional Virasoro algebra describing 
the CFT. On the other hand, it has been argued in the past that 
the asymptotic dynamics of $AdS_3$ can also be described by a
Liouville theory \refs{\henneaux, \navarro, \bbg}. Consistency 
of these two approaches therefore suggests that, in a certain sense, 
the Liouville theory effectively describes the collective field
excitations of the CFT at the boundary \refs{\martinec}.

In a recent paper \refs{\seiberg}, it has been argued that for
the Euclidean version of $AdS_3$, it is possible to write 
down a world-sheet string action. Quantization of this action
would describe the string dynamics around the background
$AdS_3$. One would thus expect to see a version of target
space boundary CFT/Liouville theory connection from the world-sheet
physics itself. One of the purpose of this note is to explore this
issue from the world-sheet point of view. In particular, 
we will perturb the action with an operator which, in turn, generates
the fluctuations of the target space boundary. The coupling constant
(which, however, is a function of boundary coordinates) 
associated with the operator, needs to satisfy certain consistency 
conditions, namely the conformal invariance condition on the 
world-sheet. Preserving conformal invariance, we then discuss the
emergence of a Liouville like field in the target space which
effectively describes the boundary fluctuations. Moreover, this 
naturally allows us to construct various explicit deformations 
of $AdS_3$ by properly choosing the world-sheet coupling.
 
An important
class of three dimensional geometry is described by the
BTZ black hole \refs{\btz, \strominger, \bss, \carlip}. Though globally
different, 
the local structure
of such black holes are that of $AdS_3$. 
In particular, the zero mass BTZ black hole is described 
locally by the $AdS_3$ metric.  
In \refs{\strominger, \ms},
these properties were exploited in order to calculate the 
entropy of the BTZ black hole by explicitly counting the
number of states. In this note, we will try to argue that
certain boundary fluctuations for these singular geometries 
correspond
to mass and angular momentum fluctuations provided the 
world-sheet coupling satisfies two differential equations
with respect to the boundary coordinates. It turns out that
these equations can be solved exactly in order to get the 
right behaviour of the world-sheet coupling. 

Before we discuss these issues, we would like to point
out that our analysis will be inherently perturbative. Thus our
results are expected to be valid when the world-sheet coupling
that corresponds to the perturbation is small.

The three dimensional gravity action 
\eqn\grav{ S_t = \int d^3x {\sqrt g}(R + {2\over l^2})}
has a classical solution which is $AdS_3$. The cosmological
constant is $\Lambda = -2$ (here and in the rest of our discussion we
set $l = 1$). The subscript 
$t$ on $S$ denotes that it is the target space action. 
The Euclidean version of $AdS_3$ metric is given by \refs{\seiberg}
\eqn\metr{ds^2 = d\phi^2 + e^{2\phi} d\gamma d\bar\gamma,}
where 
\eqn\defcor{\eqalign{& \phi = {\rm log} (X_{-1} + X_3),\cr
&\gamma = {{X_2 + i X_1}\over {X_1 + X_3}},\cr
&\bar\gamma  = {{X_2 - i X_1}\over {X_1 + X_3}}}}
with the constraint
\eqn\const{-X_{-1}^2 + X_3^2 + X_1^2 + X_2^2 = -1.}
In the above, we have taken $X_{-1} > 0$ and $X_{-1} > |X_3|$ so that the
expression for $\phi$ is well defined.
In this parametrization, the boundary of $AdS_3$ corresponds to 
 $\phi \rightarrow \infty$ and the boundary is a two dimensional sphere
parametrized by $(\gamma , \bar\gamma )$ .

The world sheet string action formulated in the 
background of the above metric and the NS two-form field $B =
e^{2\phi} d\gamma \land d\bar\gamma$ is given by \refs{\seiberg}
\eqn\action{ S = \int dzd\bar z [\partial\phi\bar\partial\phi + 
e^{2\phi}\bar\partial \gamma\partial \bar\gamma] ,}
where the string length  is also set to unity.
Notice that in the limit $\phi \rightarrow \infty$, the dominant
contribution to the path integral comes from the second term in the above
action which simply describes a two dimensional sphere.
We note that the cosmological constant in \grav ~can be thought of
coming from the kinetic term of the two-form field of string theory. 
In the sigma model language, this term corresponds to the
contribution from the $B_{\mu\nu}$ term. Thus, the action in \action 
 ~is the sum of contributions from the target space metric  $G_{\mu\nu}$
and
the $B_{\mu\nu}$ field. For consistency of string theory, we also assume
here that the action \action
~is coupled to some internal conformal field theory
in order to have total central charge 26. Except balancing the 
central charge, the internal CFT will not play much role in our
following discussion. 

The simplest way to study small fluctuations around $AdS_3$ would 
be to deform the action \action ~by perturbing it with certain 
operators. We will argue below that one such interesting  operator
which captures the effect of small boundary fluctuations is
$f(\gamma,\bar\gamma ) e^{2\phi}\bar\partial\gamma\partial \bar\gamma$.
Here we have introduced $f(\gamma,\bar\gamma )$ which is a function
of boundary coordinates. 
In order to describe a consistent string theory, which yields the target
space action \grav, the perturbed
action needs to satisfy certain constraints. 
In other words, world sheet
$\beta$-function vanishing conditions would give some restrictions on the 
function $f(\gamma,\bar\gamma )$. We will comment on this issue and its
use in our analysis only towards the end of our discussion. At present, we
use the freedom of field redifinition and gauge transformation to recast
the perturbed action in form which is simpler for our purpose.

Thus, we start with the  perturbed action of the form
\eqn\paction{ S_p = \int dzd\bar z [\partial\phi\bar\partial\phi + 
e^{2\phi}\bar\partial \gamma\partial \bar\gamma] 
+ 2 \int dz d\bar z
f(\gamma,\bar\gamma ) e^{2\phi}\bar\partial\gamma\partial 
\bar\gamma .}
Clearly, the perturbation makes sense when the coupling
parameter $f(\gamma , \bar\gamma)$ is small. So our analysis 
can only be taken seriously in the region where $f$ as a function
of $\gamma, \bar\gamma$ is small and slowly varying. The factor of 2 in
the perturbing term is taken for convenience.

As mentioned above, it is convenient to make the following 
redefinition of fields :
\eqn\fieldredef{\eqalign{ &\tilde \phi = \phi +  f(\gamma, \bar\gamma), \cr
&\tilde\gamma = \gamma + e^{- 2\tilde \phi}{{\partial} 
f\over {\partial\bar\gamma}}\cr
& {\bar{\tilde\gamma}} = \bar\gamma +
 e^{- 2\tilde \phi}{{\partial }f\over
{\partial\gamma}}.}}

With this change of variables, the action \paction
~reduces to the following form
\eqn\rpaction{
S_p = \int dzd\bar z [F_{\tilde\phi\tilde\phi}\partial \tilde\phi
\bar\partial 
\tilde\phi + F_{\tilde \gamma\tilde\gamma}\partial {\tilde\gamma}
\bar\partial{\tilde\gamma} + 
 F_{ {\bar{\tilde\gamma}}{\bar{\tilde\gamma}}}\partial {\bar
{\tilde\gamma}}
\bar\partial{\bar{\tilde\gamma}} + 
 F_{ {{\tilde\gamma}}{\bar{\tilde\gamma}}}\partial {
{\tilde\gamma}}
\bar\partial{\bar{\tilde\gamma}} +
F_{ {{\bar{\tilde\gamma}}}{{\tilde\gamma}}}\partial 
{\bar{\tilde\gamma}}
\bar\partial{{\tilde\gamma}} + D_{\tilde\phi \tilde\gamma 
{\bar{\tilde\gamma}}}],} 
where,
\eqn\fs{\eqalign{
 &F_{\tilde\phi\tilde\phi} = 1 - 4 e^{-2\tilde\phi}\partial_{\tilde\gamma}f
\partial_{\bar{\tilde\gamma}}f  ,\cr
& F_{ {{{\tilde\gamma}}}{{\tilde\gamma}}} = \partial_{\tilde\gamma}f
\partial_{{\tilde\gamma}}f - 
\partial_{\tilde\gamma}\partial_{\tilde\gamma}f , \cr
& F_{ {{\bar{\tilde\gamma}}}{\bar{\tilde\gamma}}} = 
\partial_{\bar{\tilde\gamma}}f \partial_{\bar{\tilde\gamma}}f -
\partial_{\bar{\tilde\gamma}}\partial_{\bar{\tilde\gamma}}f ,\cr
&F_{ {{{\tilde\gamma}}}{{\bar{\tilde\gamma}}}} = 
\partial_{\bar{\tilde\gamma}}f \partial_{{\tilde\gamma}}f -
\partial_{\bar{\tilde\gamma}}\partial_{{\tilde\gamma}}f ,\cr
&F_{ {{\bar{\tilde\gamma}}}{{{\tilde\gamma}}}} = 
e^{2\tilde\phi} + \partial_{\bar{\tilde\gamma}}f
\partial_{{\tilde\gamma}}f -
\partial_{\bar{\tilde\gamma}}\partial_{{\tilde\gamma}}f ,\cr
& D_{\tilde\phi \tilde\gamma
{\bar{\tilde\gamma}}} = 
\partial_{{\tilde\gamma}}f\partial\tilde\phi\bar\partial\tilde\gamma
+ \partial_{\bar{\tilde\gamma}}f\bar\partial
{{\tilde\phi}}\partial{\bar{\tilde\gamma}}\cr
&~~~~~~~~-\partial_{\bar{\tilde\gamma}}f\partial
{{\tilde\phi}}\bar\partial{\bar{\tilde\gamma}}
- \partial_{{\tilde\gamma}}f\bar \partial\tilde\phi\partial\tilde\gamma. }}
In deriving the above expressions, we have neglected the subleading
corrections
in powers of $e^{-2\tilde\phi}$. In the following, we 
remove the tildes for notational simplicity.

We now choose the function $f(\gamma,\bar\gamma)$ such that
its $\gamma$ and $\bar\gamma$ dependences decouple. Namely, we would
consider $f(\gamma,\bar\gamma) = g (\gamma) + h (\bar\gamma )$. 
Though at this stage, this choice looks ad-hoc, we will justify it
at a later stage of the paper.

Equation \rpaction~can be brought to a much simpler form
by exploiting various gauge symmetries of the target space
fields. First of all, 
we notice that $D_{\phi \gamma{\bar{\gamma}}}$
can be removed away from the action by making gauge transformations 
of the antisymmetric field $B_{\mu\nu}$.
Under gauge transformation, $B_{\mu\nu} \rightarrow B_{\mu\nu}
+ \partial_\mu \lambda_\nu - \partial_\nu \lambda_\mu$ where $\mu,\nu$
are the target space coordinates. It is easy to show that
two such  successive gauge transformations will precisely generate 
$D_{\phi \gamma{\bar{\gamma}}}$.
First we choose 
$\lambda_\phi = g(\gamma )$ and  $\lambda_\gamma = 
\lambda_{\bar\gamma} = 0$
and the second with 
$\lambda_\phi = -h(\bar\gamma )$ and $\lambda_\gamma =
\lambda_{\bar\gamma} = 0$.
In other words, 
$D_{\phi \gamma{\bar{\gamma}}}$ is a total derivative
term since
\eqn\dpg{D_{\phi \gamma{\bar{\gamma}}}
={1\over 2}[ \partial_\alpha(\epsilon^{\alpha\beta}g \partial_\beta\phi ) 
  - \partial_\alpha(\epsilon^{\alpha\beta}h \partial_\beta\phi )].}
Here, $\alpha$ and $\beta$ are world sheet coordinates. Thus we will not 
consider $D_{\phi \gamma{\bar{\gamma}}}$ further.

In general, the freedom of adding total derivative terms on the
world-sheet correspond to gauge transformations of the target
space field. One such example has just been discussed above.
We will use this freedom again to simplify \rpaction ~further.
Let us first note that
\eqn\td{{-{1\over 2}}\partial_\alpha 
(\epsilon^{\alpha\beta}h\partial_\gamma g
\partial_\beta \gamma ) = \partial_{\bar\gamma}h \partial_{\gamma}
g\partial{\bar\gamma}\bar\partial\gamma - 
 \partial_{\bar\gamma}h \partial_{\gamma}g 
\bar\partial{\bar\gamma}\partial\gamma .}
If we add this total derivative term in the action \rpaction,
the only changes appear in the $F_{\gamma\bar\gamma},
F_{{\bar\gamma}\gamma}$ components. Infact, they now have the following 
forms:
\eqn\somef{F_{\gamma{\bar\gamma}} = 0,~~~F_{{\bar\gamma}\gamma}
= e^{2\phi} + 2  \partial_{\bar\gamma}h \partial_{\gamma}g.}
As an aside, let us note that the addition of this total
derivative term can also be understood as gauge variation of 
$B_{\mu\nu}$ as before for suitable choice of $\lambda$.
Thus, near the asymptotic boundary, $\phi \rightarrow \infty$,
we can further simplify the action to bring it to the following form:
\eqn\simple{S_p = \int dzd\bar z [F_{\phi\phi}\partial\phi
\bar\partial
\phi + F_{ \gamma\gamma}\partial {\gamma}
\bar\partial{\gamma} +
 F_{ {\bar{\gamma}}{\bar{\gamma}}}\partial {\bar
{\gamma}}  
\bar\partial{\bar{\gamma}} +
F_{ {{\bar{\gamma}}}{{\gamma}}}\partial  
{\bar{\gamma}}
\bar\partial{{\gamma}}], 
}
where 
\eqn\newfs{\eqalign{
 &F_{\phi\phi} = 1 ,\cr
& F_{ {{{\gamma}}}{{\gamma}}} = \partial_{\gamma}g
\partial_{{\gamma}}g -
\partial_{\gamma}\partial_{\gamma}g , \cr
& F_{ {{\bar{\gamma}}}{\bar{\gamma}}} =
\partial_{\bar{\gamma}}h \partial_{\bar{\gamma}}h -
\partial_{\bar{\gamma}}\partial_{\bar{\gamma}}h ,\cr
&F_{ {{\bar{\gamma}}}{{{\gamma}}}} =
e^{2\phi}.}}

With such a simple form of the world sheet action, it is
now easy to read of the target space metric. This simply is
\eqn\tragetmetric{
ds^2 =  d\phi^2 + e^{2\phi}d\gamma 
d\bar\gamma + F_{\gamma\gamma}d\gamma^2 + F_{\bar\gamma
\bar \gamma}d{\bar\gamma}^2 .}
Now, following \refs{\brown, \navarro}, we can identify, up to a constant, 
$ F_{\gamma\gamma}, F_{\bar\gamma \bar\gamma}$ as 
components of stress-energy tensor in the boundary conformal field
theory and their form as given in \newfs very much suggests that the
boundary theory is a Liouville field theory. Thus the world-sheet
perturbation does infact lead to Liouville theory in the boundary of
$AdS_3$.

We know that the action \action ~has a $SL(2)$ free field representation
which is discussed in detail in \refs{\seiberg}. 
Equation \action ~can be written as 
\eqn\slaction{ S = \int dzd\bar z [\partial\phi\bar\partial\phi
+ \beta\bar\partial\gamma + \bar\beta \partial\bar\gamma
- e^{-2\phi}\beta\bar\beta ].}
Now if we use the equations of motion of $\beta$ and $\bar\beta$
in \slaction, we recover \action. Notice that at the boundary,
the contribution of the screening operator $e^{-2\phi}\beta\bar\beta$
is small and thus can be neglected. However, as we move 
from the boundary to bulk, it is this term that starts 
contributing in the correlation function. 
In the $SL(2)$ language, \paction~corresponds to changing the
coefficient of the screening operator by a slowly varying 
$\gamma, \bar\gamma$ dependent function as
\eqn\pslaction{S_p = \int dzd\bar z ( \partial\phi\bar\partial\phi
+ \beta\bar\partial\gamma + \bar\beta \partial\bar\gamma
-  [1 - 2 f(\gamma, \bar\gamma)]
e^{-2\phi}\beta\bar\beta ).}
Integrating out $\beta$ and $\bar\beta$, we get \paction. 

The reason for discussing the $SL(2)$ action is to point out
a shortcoming of our previous analysis. It
is well known that at the
quantum level, \slaction ~receives corrections. The field $\phi$,
in particular, acquires a world-sheet curvature coupling. We 
were not careful about such contribution in our previous analysis.
Thus, while valid classically, our discussion will need proper
care of quantum contributions. This certainly deserves further
investigation. 

With the above remark in mind, we now use our  results
in specific contexts.
Let us first consider the special case where $f$ is a function of
only $\gamma$ in \paction. In this case, the only contribution
in the subleading order that survives in \fs~is $F_{\gamma\gamma}$.
On the other hand, from explicit calculation, we find  that any metric of 
the 
form 
\eqn\sol{ds^2 = d\phi^2 + e^{2\phi} d\gamma d\bar\gamma + G(\gamma )
d\gamma^2}
satisfies the Einstein equation that follows form
\grav. Here $G(\gamma )$ is arbitrary function 
of $\gamma$. However, comparing with \tragetmetric, we see
that \sol~ can be identified with the deformation of $AdS_3$, if
we identify $F_{\gamma\gamma} = G(\gamma)$. Thus we get an 
equation of the form:
\eqn\deter{\partial_\gamma\partial_\gamma g -
\partial_\gamma g\partial_\gamma g + G(\gamma) = 0.}
Given a functional form of $G(\gamma)$, we can solve \deter
~in order to find $g$. Upon substitution of $g$ in \paction, 
we get action around the perturbed $AdS_3$ background. This, in turn,
gives us an well defined prescription to understand {\it small}  
fluctuations of $AdS_3$ metric by perturbing \action ~with 
suitable coupling. This coupling can be determined uniquely
by equation \deter ~or its generalisations (which we discuss
later). Turning this argument around, we see that for a particular
choice of $g$, the metric perturbation gets determined through
\deter. For example, when $g \sim \lambda_p \gamma^p$ with 
$\lambda_p$ being constant, deformation around $AdS_3$ is given by 
$G(\gamma ) \sim \lambda_p \gamma^{p-2}(p\gamma^p -1)$.
Moreover, it is easy to check that these deformations 
are due to conformal transformations on the $AdS_3$ boundary.
Under boundary conformal transformation 
the boundary term $e^{2\phi}
\partial\bar\gamma\bar\partial\gamma$ in the action \action~
gets modified. More precisely under
\eqn\conf{\gamma \rightarrow \gamma + y(\gamma ),~~~
S \rightarrow S + \partial_\gamma y e^{2\phi}\partial\bar\gamma
\bar\partial\gamma.}
If we now compare the modified action with \paction, we get 
$g = \partial_\gamma y/2$. Solving for $y$, we thus have $y 
\sim \lambda_p\gamma^{p+1}$. Hence we see that 
under $\gamma \rightarrow  \gamma +  \lambda_p\gamma^{p+1}$, 
the $AdS_3$ metric gets deformed by 
$G(\gamma ) \sim \lambda_p \gamma^{p-2}(p\gamma^p -1)$.

Exactly same arguments repeat when target space metric is of the form 
\eqn\solo{ds^2 = d\phi^2 + e^{2\phi} d\gamma d\bar\gamma + H(\bar\gamma )
d{\bar\gamma}^2.}
In this case, the relevant coupling is of course $h(\bar\gamma )$ and 
the equation analogous to \deter~is
\eqn\detero{\partial_{\bar\gamma}\partial_{\bar\gamma} h -
\partial_{\bar\gamma} h\partial_{\bar\gamma} h + H(\bar\gamma) = 0.}

The above analysis can be easily generalized for metric deformations
that are both functions of $\gamma$ and $\bar\gamma$. Instead
of being general, let us consider a specific case of this sort.
A nice example is the BTZ metric which is a solution of \grav. In
Schwarzschild-like 
coordinate \refs{\carlip}, the metric is 
\eqn\btz{ds^2 = -N^2 dt^2 + N^{-2} dr^2 + (d\theta + N^\theta dt)^2,}
where 
\eqn\per{N^2 = (-M + r^2 + {J^2\over {4r^2}}),~~~N^\theta = {J\over
{2r^2}}.}
Here, $M$ and $J$ correspond to mass and angular momentum parameters of
the black hole respectively. In the extremal limits $J = \pm M$. 
The Euclidean version of the metric is related to the coordinates
$\phi, \gamma, \bar\gamma$ as \refs{\seiberg}
\eqn\ourco{\eqalign{&r = e^\phi{\sqrt{\gamma\bar\gamma}},\cr
&\theta = {1\over {2i}}{\rm log}{\gamma\over{\bar\gamma}},\cr
&\tau = \phi - {1\over 2} {\rm log}(e^{2\phi}\gamma\bar\gamma ).}}
Here $\tau$ is the analytically continued $t$. In this 
coordinate system the metric \btz~for large $\phi$ takes the form
\eqn\btzap{ds^2 = d\phi^2 + e^{2\phi} d\gamma d\bar\gamma
+ ({{-1+J-M}\over {4\gamma^2}}) d\gamma^2 + ({{-1-J-M}\over 
{4{\bar\gamma}^2}}) d{\bar\gamma}^2.}
This is easy to check since for large $\phi$, \ourco~ simplifies to
$e^\phi \sim re^\tau, \gamma \sim e^{-\tau + i\theta}$ and 
$\bar\gamma \sim e^{-\tau - i\theta}$. We note that for $J = 0,
M = -1$, the metric reduces to $AdS_3$.
Now, comparing \btzap~with \tragetmetric, we get
\eqn\comp{\eqalign{
&\partial_\gamma \partial_\gamma g - \partial_\gamma g \partial_\gamma g
+ ({{-1+J-M}\over {4\gamma^2}}) = 0,\cr
&\partial_{\bar\gamma} \partial_{\bar\gamma} h - \partial_{\bar\gamma} h
\partial_{\bar\gamma} h
+ ({{-1-J-M}\over {4{\bar\gamma}^2}}) = 0.}}
The equations in 
\comp~can be easily solved. The solutions are:
\eqn\compsol{\eqalign{&g = -{\rm log}[{\sqrt \gamma} - {\rm 
cos}({{\sqrt{M-J}} \over 2}{\rm log}\gamma )],\cr
&h = -{\rm log}[{\sqrt {\bar\gamma}} - {\rm cos}({{\sqrt{M+J}}
\over 2}{\rm log}\bar\gamma )].}}
In extremal limits, the solutions simplify considerably. For example
when $M = J$, we have:
\eqn\extrem{\eqalign{&g =  -{\rm log}[{\sqrt \gamma} -1],\cr 
&h = -{\rm log}[{\sqrt {\bar\gamma}} - {\rm cos}({{\sqrt{M\over 2}}
}{\rm log}\bar\gamma )].}}
Even if we have the exact solutions for $g$ and $h$,
we would like to emphasize again that our analysis 
above is an approximate one and expected to be valid  only in the region 
where $g,h$ are small.

Finally, we would like to end this note with a comment on the
conformal invariance of the perturbed action \paction. The world-sheet
action is conformal invariant when the $\beta$-functions vanish.
For our case, these equations are
\eqn\bvanish{\eqalign{&R_{\mu\nu} - {1\over 2}g_{\mu\nu}R -
{1\over 4}[ H_{\mu\lambda\rho} {H_{\nu}}^{\lambda\rho} - {1\over 
6}g_{\mu\nu}H^2 ] = 0,\cr
&\nabla_\lambda {H^{\lambda}}_{\mu\nu} = 0.}}
Calculating various terms from \paction, we get a constraint on $f(\gamma,
\bar\gamma)$. This is
\eqn\bcons{ (1+2f)\partial_\gamma \partial_{\bar\gamma} f
- 2 \partial_\gamma f  \partial_{\bar\gamma} f = 0.}
Defining $Y = {\rm log} (1+2f)$, above equation reduces to
\eqn\yeq{\partial_\gamma \partial_{\bar\gamma} Y = 0.}
So we see that in terms of the $Y$, conformal invariance requires
$Y$ to satisfy free field equation. Now for small enough $f$, $Y \sim 2f$.
Thus, we can approximate $f$ to satisfy free field equation, leading to
$f (\gamma,\bar\gamma ) = g(\gamma ) + h (\bar\gamma )$. This is what
has been used earlier in our discussion.

To conclude, we have demonstrated that the three dimensional gravity
corresponding to the $AdS_3$ space time with a negative cosmological
constant can be obtained from a string theory formulated in the background
of this spacetime metric and a NS two form field. The NS two form field in
the world-sheet theory is such that its contribution to the target space
exactly
gives the cosmological constant. Then a consistent perturbation can be
made in the world-sheet theory which can correspond to the boundary
fluctuations in the $AdS_3$ space time. This fluctuation can be
interpreted
as boundary diffeomorphism and it leads to a Liouville theory living on
the boundary. Using the
fact that the local structure of a BTZ black hole corresponds to $AdS_3$
space time, we could demonstrate the relationship between the pertubation
in
the world-sheet theory and the physical quantities like mass and angular
momentum of BTZ black hole.

\bigskip
\noindent{\bf Acknowledgements}: 

\noindent It is a pleasure to thank Sumit Das, Cesar Gomez
and Ashoke Sen for sharing their insights with us. We also thank Patrick
Meessen for discussion.
The work of SM is supported by Ministerio de Educaci\'on y
Cultura of Spain and also through the grant CICYT-AEN 97-1678.
We also like to acknowledge the hospitality of ICTP and SISSA during
the initial stages of the work.
\vfill\eject
\listrefs
\bye